\documentclass[a4paper,8pt]{article}
\usepackage{amsmath,amssymb}

\usepackage{graphicx}
\usepackage{epsfig,authblk}
\usepackage{bm,fullpage}

\usepackage{array,textcomp}
\usepackage{multirow}
\usepackage{hhline}

\title{\bf Universality in Bacterial Colonies}
\author[1]{Juan A. Bonachela\footnote{Corresponding author: jabo@princeton.edu}}\author[1,2]{Carey D. Nadell}\author[3]{Jo\~ao B. Xavier}\author[1]{Simon A. Levin}
\affil[1]{Department of Ecology and Evolutionary Biology, Princeton University, Princeton, NJ, 08544-1003, USA}
\affil[2]{Department of Molecular Biology, Princeton University, Princeton, NJ, 08544-1033, USA}
\affil[3]{Program in Computational Biology, Memorial Sloan-Kettering Cancer Center, 1275 York Avenue, Box 460, New York, NY, 10065, USA}
\date{}

\begin{document}
\maketitle
\begin{small}\abstract{The emergent spatial patterns generated by growing bacterial colonies have been the focus of intense study in physics during the last twenty years. Both experimental and theoretical investigations have made possible a clear qualitative picture of the different structures that such colonies can exhibit, depending on the medium on which they are growing. However, there are relatively few quantitative descriptions of these patterns. In this paper, we use a mechanistically detailed simulation framework to measure the scaling exponents associated with the advancing fronts of bacterial colonies on hard agar substrata, aiming to discern the universality class to which the system belongs. We show that the universal behavior exhibited by the colonies can be much richer than previously reported, and we propose the possibility of up to four different sub-phases within the medium-to-high nutrient concentration regime. We hypothesize that the quenched disorder that characterizes one of these sub-phases is an emergent property of the growth and division of bacteria competing for limited space and nutrients.}\end{small}

\section*{Introduction}
The use of statistical physics to unravel biological problems has yielded many novel insights into the living world. Numerous examples can be cited, including the application of statistical mechanics to ecological or metabolic networks \cite{ex_networks1,ex_networks2}, or the use of phase transition theory and correlation functions to understand the behavior of animal collectives \cite{ex_PT1,ex_PT2}. One of the most remarkable instances of marriage between physics and biology concerns the emergent spatial patterns produced by growing bacterial colonies.

\vspace{0.25cm}
\noindent
Beginning with a strip or drop of bacterial culture inoculated on an agar plate, the growth and division of cells over time results in an emergent colony pattern that varies with nutrient and agar concentration in the underlying medium. This system presents many different biological and physical questions whose answers require the application of different conceptual approaches. For instance, we can analyze the influence of cell group structure on the evolution of cooperative or competitive secretion phenotypes, measuring the reproductive fitness of idealized strains \cite{Carey_paper,CModel_ex0}. On the other hand, if we are interested in a morphological characterization of the colony's propagating front, different physical properties of the system must be analyzed.

\subsubsection*{{\it The Morphological Phase Diagram of Bacterial Colonies}}

\begin{figure}[t]
\begin{center}
\includegraphics[width=8cm]{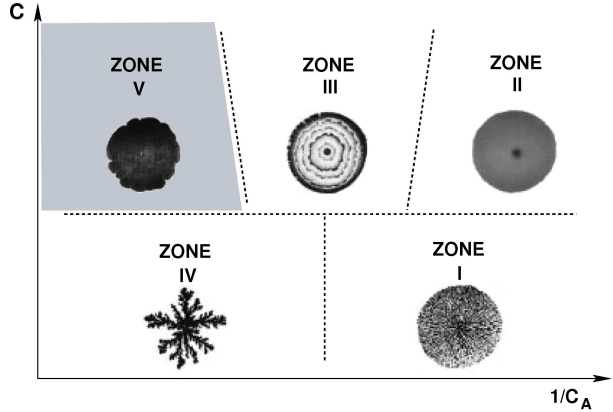}
\caption{\footnotesize{Sketch of the morphological phase diagram shown by bacterial spatial patterns (adapted with permission from \cite{rings1}; copyrighted by the Physical Society of Japan). Our work will be focused on the shaded zone, called here the {\it rough regime}.}}
\label{orig_phase_diagram}
\end{center}
\end{figure}

More than two decades ago, Matsuyama {\it et al.} \cite{fractal_bac1} and Fujikawa and Matsushita \cite{fractal_bac2} showed that bacterial colony patterns obtained in the laboratory can be fractal objects \cite{barabasi_book}. The properties of the emergent fractal depend mainly on two factors: {\it i)} nutrient concentration, $C$, strongly influences cell growth rate, and {\it ii)} agar concentration, $C_{A}$, controls cell mobility by altering the hardness of the substratum on which the colony is expanding. In the absence of special forms of collective cell motility \cite{Cmodel_new}, bacterial colony patterns may be classified with a two-dimensional phase diagram. Using {\it Bacillus subtilis}, Fujikawa, Matsushita and collaborators \cite{fractal_bac2,bac_phases0,bac_phases} defined a phase diagram that comprises five regions with different qualitative behaviors resulting from the interaction between biological and physical factors (see Fig.\ref{orig_phase_diagram}):

\begin{itemize}
\item[\texttwelveudash] At low agar concentration, cells can exhibit active flagellar motility. At low nutrient concentration (zone I), individuals tend to deplete local nutrient supplies rapidly and disperse in search of zones with better growth conditions. The resulting colonies consist of many long, thin branches (the so-called dense branched morphology, DBM) \cite{ben-jacob}. When nutrient concentration is increased, cells grow to higher densities with increased resource availability along the advancing front; the resulting pattern is compact and smoothly circular (zone II).
\item[\texttwelveudash] At medium agar concentration and relatively high nutrient availability, colonies consist of concentric rings that result from periodic alternation between static and motile stages with positive growth rates (zone III) \cite{rings0,rings1}.
\item[\texttwelveudash] At high agar concentration, the substratum becomes hard and dry enough that cells cannot move by active means. Thus, low nutrient concentration results in patterns similar to the DBM, but with fewer, thicker branches due to limited cell movement \cite{bac_phases0,bac_phases,ben-jacob}. These patterns (zone IV) resemble those created by diffusion-limited aggregation (DLA) processes \cite{DLA}, with a very similar fractal dimension \cite{fractal_bac1,fractal_bac2}. For medium-to-high nutrient levels, the colony patterns are compact again, but with more irregular fronts than those obtained with low agar concentration (zone V) \cite{bac_phases,vicsek1}.
\end{itemize}

\vspace{0.25cm}
Many different models have managed to reproduce parts or all of the morphological phase diagram by adding new biological ingredients, such as cooperativity by means of chemotactic signaling \cite{ben-jacob}, or new physical ingredients, such as instabilities induced by fluctuations during cell diffusion \cite{levine}. Matsushita {\it et al.} \cite{bac_model0} introduced a three-equation reaction-diffusion model capable of reproducing, at least phenomenologically, all regions of the phase diagram described above.

\vspace{0.25cm}
Thanks to this considerable amount of both experimental and theoretical work, the qualitative picture of the bacterial colony pattern phase diagram is well understood.

\subsubsection*{{\it Quantitative Description}}

Bacterial colony patterns are excellent candidates for analysis via propagating-front theory, due to a correspondence between the biology of bacterial growth and the physical aspects of nonliving fronts studied in statistical physics. Let us consider high agar concentrations (that is, increased surface hardness and minimized bacterial mobility) and an initially flat inoculum. As bacteria grow and divide, they passively shove each other, which causes the colony front to advance in space. Let $h(x,t)$ be the distance, at time $t$, between the top of the cell group and the basal substratum at its linear coordinate $x$ (see Fig.\ref{check}). Importantly, not all the cells can grow and divide. Only those that can access nutrients, namely those forming an active layer at the advancing front, can do so (see inset in Fig.\ref{check}). The depth of this active layer depends on nutrient penetration into the group. The identification of growth and shoving as the main driving forces allows us to draw an analogy with non-living propagating fronts or driving interfaces from statistical physics, like those formed by fire on a burning sheet of paper or water absorbed into a porous medium \cite{barabasi_book}.

\vspace{0.25cm}
The main observable measured in the study of these fronts is the variance in surface height, also termed the interface width or roughness (a measurement of the lateral correlations at the interface). It is defined as:

\begin{equation}
W(L,t)=\left<\overline{\left[h(x,t)-\overline{h}\right]^{2}}\right>^{1/2},
\label{width} 
\end{equation}

\noindent
where the overline represents spatial averages, $<.>$ the average over runs of the experiment, and $L$ is the inoculation strip length (system size, hereafter). A family of different $2$-dimensional colony patterns may arise, depending on the environmental conditions. If the colony pattern is fractal or compact, the front can be treated {\it a priori} as a self affine object, and the so-called Family-Vicsek scaling may be assumed \cite{FV}:

\begin{equation}
W(L,t)\sim\left\lbrace
\begin{array}{ll}
t^{\beta}&\textnormal{ for } t<t_{x}\\
& \\
L^{\alpha}&\textnormal{ for } t>t_{x}
\end{array}
\right..
\label{family} 
\end{equation}

\noindent
That is, for short times the colony roughness follows a power law behavior defined by the growth exponent, $\beta$; at time $t=t_{x}$, it saturates to a size-dependent value  $W=W_{st}(L)$ whose finite-size scaling defines the roughness exponent, $\alpha$. From Eq.(\ref{family}) we can deduce a power-law dependence of the saturation time on size:

\begin{equation}
t_{x}\sim L^{\alpha/\beta},
\label{zeta} 
\end{equation}

\noindent
where $z=\alpha/\beta$ is the so-called dynamic exponent. Thus, two independent exponents determine the behavior of the interface.

\vspace{0.25cm}                                                                                         
Vicsek {\it et al.} \cite{vicsek1} performed the first experimental measurement of one of these scaling exponents. They obtained a non-trivial value for the experimental roughness exponent, $\alpha_{exp}$, associated with the growth of {\it Escherichia coli} and {\it B. subtilis}. Some time later, similar values were independently obtained using {\it B. subtilis} in the rough regime \cite{bac_phases}. 

\vspace{0.25cm}
\noindent
In the experimental measurements reported in the literature described above, the $\alpha_{exp}$ exponent was measured by choosing different parts of a growing colony as replicates at successive time points of their saturation regimes. At each of these time points, several boxes of lateral size $l$ were used for the measurement of finite size effects over the roughness of the interface \cite{bac_phases,vicsek1}. Let us briefly analyze this observable, which we designate $w_{exp}$. If we define the ``local roughness'' of the front as \cite{local_w}:

\begin{equation}
w(l,t)=\left< \langle\left[ h(x,t)-\langle h\rangle_{l}\right]^{2}\rangle_{l}\right>^{1/2},
\label{wl}
\end{equation}

\noindent
where $<.>_{l}$ represents a spatial average over boxes of size $l$, the behavior of this local observable resembles that of the global variable, but with different exponents:

\begin{equation}
w(l,t)\sim\left\lbrace
\begin{array}{ll}
t^{\beta}&\textnormal{ for } t<t'_{x}\\
& \\
l^{\alpha_{loc}}&\textnormal{ for } t>t'_{x}
\end{array}
\right..
\label{wl_scaling} 
\end{equation}

\noindent
Systems in which $\alpha\neq\alpha_{loc}$ show what is called {\it anomalous roughening}, while regular self-affine interfaces fulfill the standard scaling for the roughness $\alpha=\alpha_{loc}$ \cite{anom_rough}. For $l\sim a$, where $a$ is the grid spacing (or, in this case, the cell diameter), $w(l\sim a,t)$ exhibits power-law behavior with positive exponent for cases of anomalous roughening, while it is constant or decreases with time for standard self-affine interfaces \cite{local_w}. For these standard self-affine interfaces, the finite-size scaling of this local roughness provides a good measurement of the global, universal exponent $\alpha$. For interfaces showing anomalous roughening, however, the values of local exponents are not necessarily universal \cite{kappa}, and we therefore cannot infer the universal behavior of the system only by measuring them. In both cases, as $l\rightarrow L$, $w(l,t)$ converges to $W(L,t)$.

\vspace{0.25cm}
\noindent
The observable measured in experiments, $w_{exp}$, is the result of performing this local finite-size scaling by averaging over multiple time points of the same run, using different parts of the same system as independent replicates. Given enough separation between time and spatial sampling points, $w_{exp}=w(l,t)$ and, thus, $\alpha_{exp}=\alpha_{loc}$. In the experimental work mentioned above, measurements of this $w_{exp}$ performed in zone V yielded a scale-free behavior of the observable, with an exponent in the range $\alpha_{exp}=0.74-0.78$ with a standard error of up to $0.07$ \cite{bac_phases,vicsek1}. This result is supported by several theoretical models \cite{bac_model2,qKPZ_MSM} (see table 1).

\begin{table}[t]
\begin{center}
\begin{tabular}{|c||c|c|c|c|c|c|}
\hline
&$\beta$&$\alpha$&$z$&$\alpha_{loc}$&$\alpha_{exp}$&\begin{footnotesize}STANDARD\ SELF-AFF.\end{footnotesize}\\
\hline
KPZ \cite{KPZ}&$1/3$&$1/2$&$3/2$&$1/2$&$-$&\checkmark\\
\hline
qKPZ\cite{qKPZ1} &$3/5$&$3/4$&$5/4$&$3/4$&$-$&\checkmark\\
\hline
qKPZ ($F>F_{c}$)&$0.62(3)$&$0.75(5)$&$1.19(5)$&$0.68(5)^{*}$&$0.72(5); 0.68(5)$&\checkmark\\
\hline
\hline
Experiments \cite{bac_phases,vicsek1}&$-$&$-$&$-$&$-$&$0.74-0.78(7)$&$-$\\
\hline
Theoretical Model&$0.65$ \cite{qKPZ_MSM}&$0.80$ \cite{qKPZ_MSM}&$1.15$ \cite{qKPZ_MSM}&$-$&$0.74$ \cite{bac_model2}&$-$\\
\hline
\hline
Present Work&$0.61(5)$&$0.68(5)^{*}$&$1.11(17)^{*}$&$0.68(5)$&$0.67(5)$&\checkmark\\
\hline
\end{tabular}
\end{center}
\label{table}
\caption{Value of the exponents for the different universality classes, experiments and numerical simulations mentioned in the text. The values with an asterisk have been calculated using Eq.(\ref{zeta}) or by using the standard self-affinity feature of $w(l,t)$. The two values for $\alpha_{exp}$ for the moving qKPZ correspond to measurements using large $L$ (left) and a system size equivalent to those used in our simulation framework (right).}
\end{table}

\vspace{0.25cm}
Thus, with simple concepts and observables commonly used in statistical physics, one may characterize the non-trivial morphology of bacterial colony fronts and quantitatively compare the colony patterns produced by different strains of bacteria. Similar values for the exponents measured in different bacterial strains indicate that, despite the varying microscopic details and interactions specific to each strain, they share the same basic biological and physical ingredients at a larger scale of observation. That is, they show the same {\it universal behavior}.

\vspace{0.25cm}
Much less attention has been paid to the universal behavior exhibited by bacterial colony patterns. Attending to the expected symmetries in the colony patterns (i.e. the different terms of the equations in the theoretical models), the current consensus is that the branches of the low nutrient regimes are the result of a DLA process (zone IV) or DBM (zone I). The concentric-ring pattern (zone III) is treated as a special case, and the flat circular pattern (zone II) as a trivial one \cite{note0}. However, the behavior of the compact pattern observed in phase V (the rough regime, from now on) remains contentious. This is mainly because the exponent associated with the only measured observable, the roughness exponent, is insufficient to be conclusive in its classification. With no other observable measured, we cannot be sure if, for instance, the interface shows anomalous roughening, in which case $\alpha_{exp}$ would not represent the global roughness of the front. In that case, the measurement of the local exponent would not provide any information about the universal behavior of the surface (see above).

\vspace{0.25cm}
\noindent
Here, we will attempt to fill this gap by studying bacterial colony fronts from the point of view of universality. By focusing on the rough regime (zone V), we will show that the behavior of the propagating surfaces can be much richer than that described in the cited experimental work, prompting a modification of the phase diagram portrayed above. Also, we will provide the whole set of exponents defined in Eqs.(\ref{family})-(\ref{zeta}) supporting our classification, and determine which are the relevant biological and physical ingredients giving rise to such values for the exponents.

\section*{A Proxy for the Experiments}
To measure experimentally the proposed set of exponents, we must be able to reproduce the environmental conditions  belonging to the regime under consideration, namely the rough regime, for each replicate. Here, cells are not actively motile, so growth and passive shoving are the sole mechanisms that induce colony front propagation.

\vspace{0.25cm}
To characterize the behavior of the bacterial colony fronts, an ideal experiment should use independent replicates to measure the exponents presented in Eq.(\ref{family})-Eq.(\ref{wl_scaling}). An exhaustive sweep of the nutrient concentration values with sufficient replication to obtain good statistics is highly time-consuming, and though we are performing such experimental work (to be reported elsewhere \cite{next_bac}), here we will use a mechanistically detailed simulation framework. Our framework is derived from the latest generation of agent-based models developed by chemical engineers to predict the structure and metabolic activity of bacterial communities \cite{CModel_paper}. The model has already been used to study colony structure patterns in the high agar concentration regime (zones IV and V), with excellent experimental support \cite{CModel_paper,ex_proxy1,ex_proxy2}. With this simulation framework, we can monitor the state of every cell and its interaction with the nutrient field and the rest of the group. Also, we have total control over environmental conditions and individual traits, including maximum growth rate and the ability to absorb nutrients \cite{note2}. The tractability offered by these individual-based simulations allows us to perform a finer exploration of the space of parameters that contribute to colony pattern formation. This will be useful not only for the goals of this paper, but also for guiding our ongoing and future work with laboratory experiments.

\vspace{0.25cm}
\noindent
Briefly, our framework comprises two overlapping layers, one of which is divided into a grid that tracks local nutrient concentration, and the other of which monitors bacterial cells that are implemented off-lattice, i.e. as rigid circles in a continuous space. We consider the growth of a generic bacterial strain whose physiological traits are parametrized to those of {\it E. coli}. Cells grow according to local nutrient concentration, and once they reach a maximum radius of $0.75\;\mu m$, they divide into two daughter cells. A shoving algorithm is used to prevent overlapping between cells as they grow and divide. The model assumes a bulk fluid in which nutrient concentration is held constant, and implements a boundary layer in which nutrient concentration profiles are calculated by solving the following reaction-diffusion equation to equilibrium at each time step:

\begin{equation}
\dfrac{\partial C_{loc}}{\partial t}=D\nabla^{2}C_{loc}-\dfrac{1}{Y}\mu,
\label{nutrient_eq}
\end{equation}

\noindent
where $C_{loc}$ is the local nutrient concentration, $D$ is the nutrient diffusivity, $Y$ is a metabolic yield coefficient, and $\mu$ is a Monod-type bacterial growth rate expression \cite{note2}. In calculating nutrient concentration profiles this way, we follow the common assumption that reaction-diffusion is much faster than bacterial growth and metabolism. More details on the simulation framework can be found in \cite{CModel_paper}. 

\begin{figure}[t]
\begin{center}
\includegraphics[width=7cm]{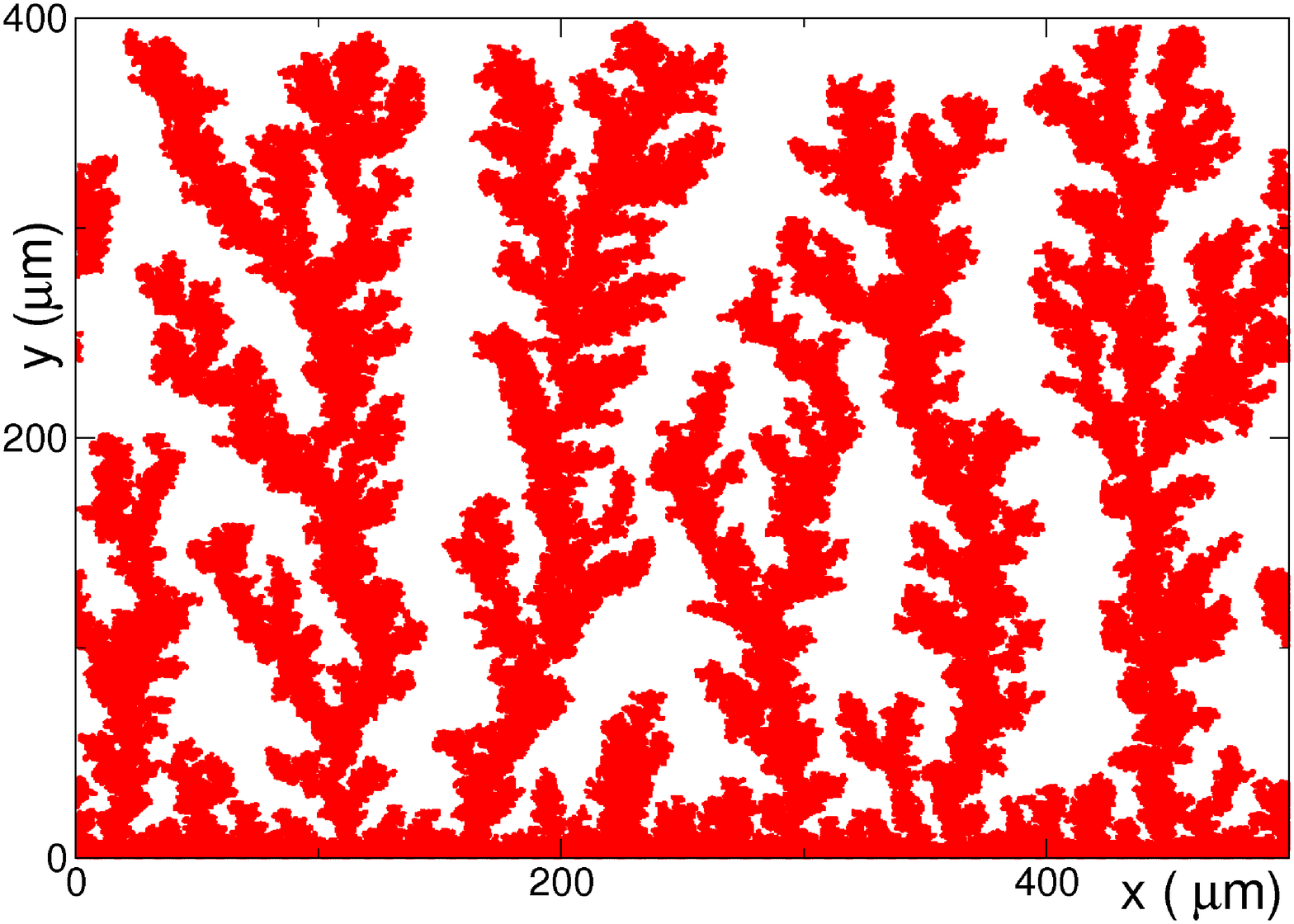}\hspace*{1cm}
\includegraphics[width=7cm]{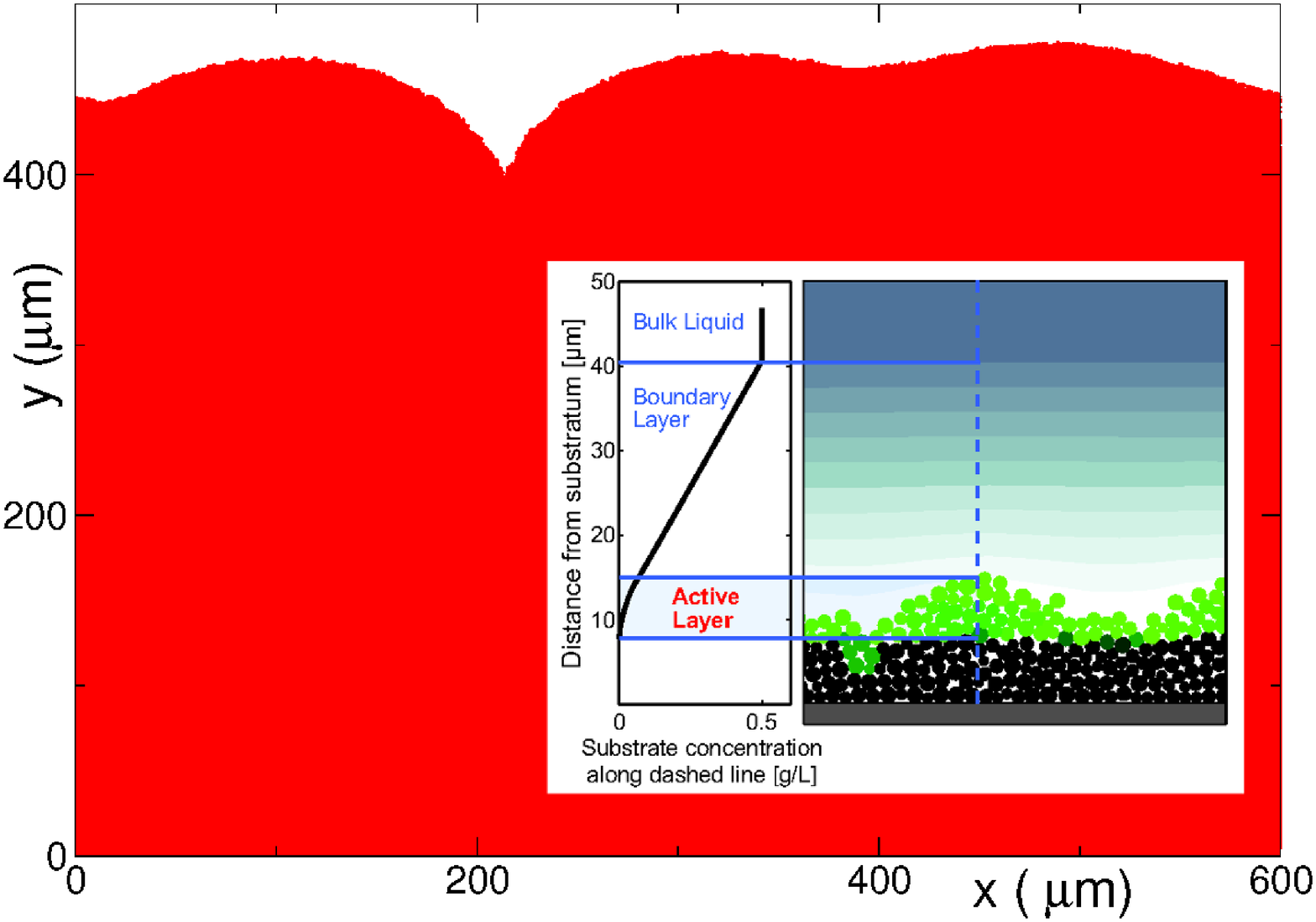}
\caption{\footnotesize{Patterns resulting from the growth of the colony starting from an initial inoculum of $L=500\;\mu m$ with periodic boundary conditions. Nutrient concentrations are $C=0.05\;g/l$ (left) and $C=3\;g/l$ (right). Inset: Illustration of the active layer (green circles) and how it distorts the bulk liquid nutrient field, creating a boundary layer; inactive cells are represented in black; reproduced from \cite{Carey_paper}.}}
\label{check}
\end{center}
\end{figure} 

\vspace{0.25cm}
Fixing all other parameters, we will vary the bulk nutrient concentration, $C$, and system size, $L$, to measure the value of the exponents defined in Eqs.(\ref{family})-(\ref{wl_scaling}). The system size will range from $L=300\;\mu m$ to $L=600\;\mu m$, covering the maximum window sizes used in experimental measurements (up to $400\;\mu m$ in \cite{bac_phases}). In addition, we will monitor $w(l\sim a,t)$, to discern whether the system shows anomalous roughening (see above). Due to technical reasons, the duration of the experiment is limited to the time needed by the colony to reach $h_{max}=L$, which depends mainly on nutrient concentration.

\vspace{0.25cm}
We begin by checking that our model generates a qualitatively similar behavior to that found in the high-agar concentration experiments. Figure \ref{check} shows that low (left) and medium (right) bulk nutrient concentration result in colony patterns similar to those of zones IV and V, respectively, with thick sparse DLA-like branches in the former case and a compact rough pattern in the latter \cite{bac_phases,vicsek1}.

\begin{figure}[t]
\begin{center}
\includegraphics[width=7cm]{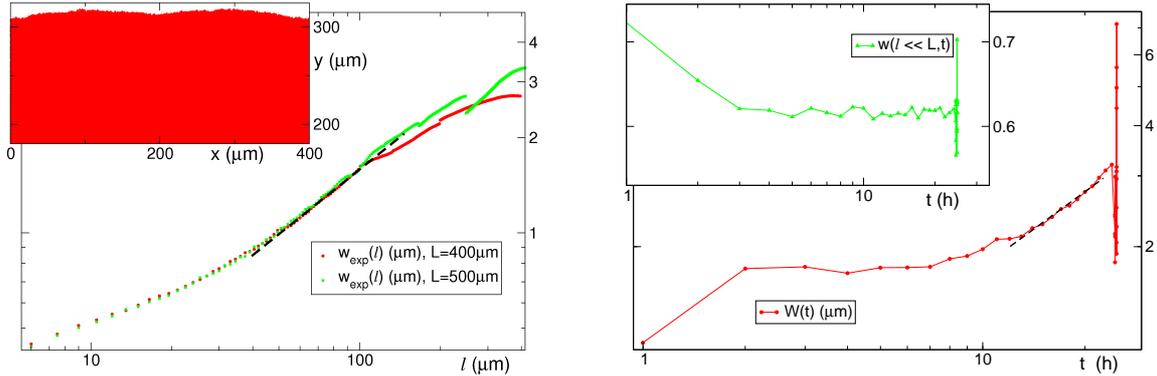}\hspace*{1cm}
\includegraphics[width=7cm]{figures/W_vs_t_delta4.5_L600_w_inset.eps}
\caption{\footnotesize{Left: Plot of the $w_{exp}(l)$ function for two different system sizes; after an initial transient, they both converge to a power law with exponent $\alpha_{exp}=0.67(5)$. The colony pattern is compact and rough (inset). Right: Dynamic behavior of the $W(t)$ function, showing a power law with an exponent $\beta=0.61(5)$ for more than $10h$ of colony growth. Right, inset: Behavior of $w(t)$; the zero slope of the curve indicates that $\alpha=\alpha_{loc}=\alpha_{exp}$.}}
\label{carey_exp}
\end{center}
\end{figure}

\vspace{0.25cm}
Now, we focus our attention on a quantitative characterization of the rough regime. For one selected nutrient concentration, $C$, belonging to this regime, we first measured the exponent $\alpha_{exp}$. One example of the $w_{exp}$ curve can be seen in the left panel of Fig.\ref{carey_exp} (red points), where an initial strip of $L=400\;\mu m$ wide was grown with nutrient concentration $C=3.8\;g/l$ (the resulting colony pattern is shown in the inset of the figure). Choosing this size, a saturating regime was found for the duration of the simulation, at which we measured $\alpha_{exp}=0.67(5)$. For the larger $l$, as can be expected, $w_{exp}$ saturates due to the approach of $w(l,t)$ to $W(L,t)$. This result is very robust, for it is shared by the whole range of compact, rough patterns tested ($\alpha_{exp}=0.62(5)-0.75(5)$). The values of the exponents measured fall into or close to the interval reported in experiments, within error bars (see table \ref{table}). 

\vspace{0.25cm}
\noindent                                                                                                                                                                                                                                                                                                                                                                                                                                                                                                                                                                                                                               
We also measured, for $L=400\;\mu m$, the local roughening exponent, obtaining $\alpha_{local}=0.68(5)$. The temporal evolution of $w(l\ll L,t)$ (see right inset, Fig.\ref{carey_exp}) shows a constant behavior, indicating that $\alpha=\alpha_{loc}=\alpha_{exp}$ for the colony. The growth exponent was measured for the largest system size ($L=600\;\mu m$), giving a value of $\beta=0.61(5)$. Using Eq.(\ref{zeta}), the derived dynamic exponent is $z=1.11(17)$. By collapsing the curves for different system sizes ranging from $L=300\;\mu m$ to $L=600\;\mu m$ \cite{note}, we independently confirmed these measurements. These exponents are in accordance with those found in \cite{qKPZ_MSM} with a reaction-diffusion model similar to that introduced in \cite{bac_model0} (see above). We now turn our attention to determining the universality class to which the system belongs.

\subsubsection*{{\it A Universality Class for the Compact, Rough Patterns}}

The Eden model is the paradigm of toy models representing growing cell colonies \cite{Eden,Eden2,Eden_KPZ}. In one of its versions, a site along the colony front is randomly chosen each time step and one new cell is placed at any of its free surrounding locations \cite{Eden2}. The basic ingredients of this model are {\it i)} a source of stochasticity working each time step, i.e. an annealed noise, due to the random choice of the site at which to deposit a new cell, and {\it ii)} lateral growth, generated by the possibility of choosing a neighboring site for the new cell. These two main ingredients are described by the mesoscopic Langevin equation known as the Kardar-Parisi-Zhang equation \cite{KPZ}:

\begin{equation}
\partial_{t}h(x,t)=\lambda(\nabla h)^{2}+D\nabla^2h+F+\sigma\eta(x,t),
\label{KPZ_eq} 
\end{equation}

\noindent
where $D$, $\lambda$, $F$ and $\sigma$ are constants, and $\eta$, an uncorrelated (white) noise. The first term in the r.h.s. represents lateral growth and the second term represents an effective diffusion generated by it. The exponents characterizing this universality class have been calculated not only numerically, but also analytically; in $1+1$ ($x$ and $h$) dimensions, they are:  $\alpha=\alpha_{loc}=1/2$, $\beta=1/3$, $z=3/2$ \cite{barabasi_book}. 

\vspace{0.25cm}
\noindent
Due to the similarities between the Eden model patterns and bacterial colonies, KPZ has been suggested as the universality class that describes colony growth. Most of the theoretical models for bacterial growth require a non-linear diffusivity function for the active cells to generate compact patterns similar to those of phase V. The simplest of these functions, in which diffusion is proportional to the local density of active cells, develops eventually the lateral growth and effective diffusion associated with the growth of the front in the KPZ equation, Eq.(\ref{KPZ_eq}) \cite{note4}.

\vspace{0.25cm}
However, the large $\alpha$ exponent found in experiments does not match that of the KPZ class (see table 1). It represents stronger lateral correlations, and has been argued to be the result of nonlocal interaction between the moving cells, or a consequence of the rod shape of the bacteria used in experiments ({\it E. coli} and {\it B. subtilis}). In the rough regime, these particular strains form folding chains lying very close together, causing lateral correlations to develop more quickly \cite{bac_phases}. Such a large $\alpha$ exponent value is also found in systems with different stochastic terms, like non-Gaussian thermal noises, or {\it quenched} noises \cite{barabasi_book,meakin_rev}.

\vspace{0.25cm}
\noindent
If we replace the thermal noise of Eq.(\ref{KPZ_eq}) by a quenched noise, $\eta(x,h)$, the interface changes its universal behavior. For $F<F_{c}$, the advance of the surface is hindered and eventually prevented by pinning sites, represented by the quenched noise. For $F>F_{c}$, the interface is able to overcome the pinning places and grows indefinitely. At $F=F_{c}$, the interface undergoes a second order phase transition between the pinned (non-moving) phase and the depinned (moving) phase, characterized by the exponents $\alpha=\alpha_{loc}=0.63(3)$, $\beta=0.67(5)$, $z=1.01(1)$ \cite{qKPZ2} (or $\alpha=3/4$ $\beta=3/5$, $z=5/4$, analytically deduced in \cite{qKPZ1}), which define the so-called {\it quenched} KPZ class (qKPZ). This scaling is retained in the moving phase ($\alpha=\alpha_{loc}=0.75(5)$, $\beta=0.62(3)$, $z=1.18(5)$, own measurements). Note the large value for $\alpha$, close to those observed in the experiments and numerical simulations of the theoretical models (see table \ref{table}). Although different sources of quenchedness – such as heterogeneities of the agar substratum \cite{barabasi_book,bac_model2} – have been proposed to influence expanding bacterial colonies, the exact origin for the large $\alpha$ exponent is still controversial.

\vspace{0.25cm}
Interestingly, our mechanistic model produces a roughness exponent similar to those reported in real experiments, but none of the proposed explanations for this exponent are present: the cells are circular, shoving events move them randomly very short distances, and there is no possibility for substratum heterogeneity in our framework. Moreover, our measurements agree with the set of theoretical exponents for the qKPZ class. Cell-cell shoving events generate the short-range lateral growth and diffusion terms and the thermal noise but, what is the source of quenchedness?

\begin{figure}[t]
\begin{center}
\includegraphics[width=7cm]{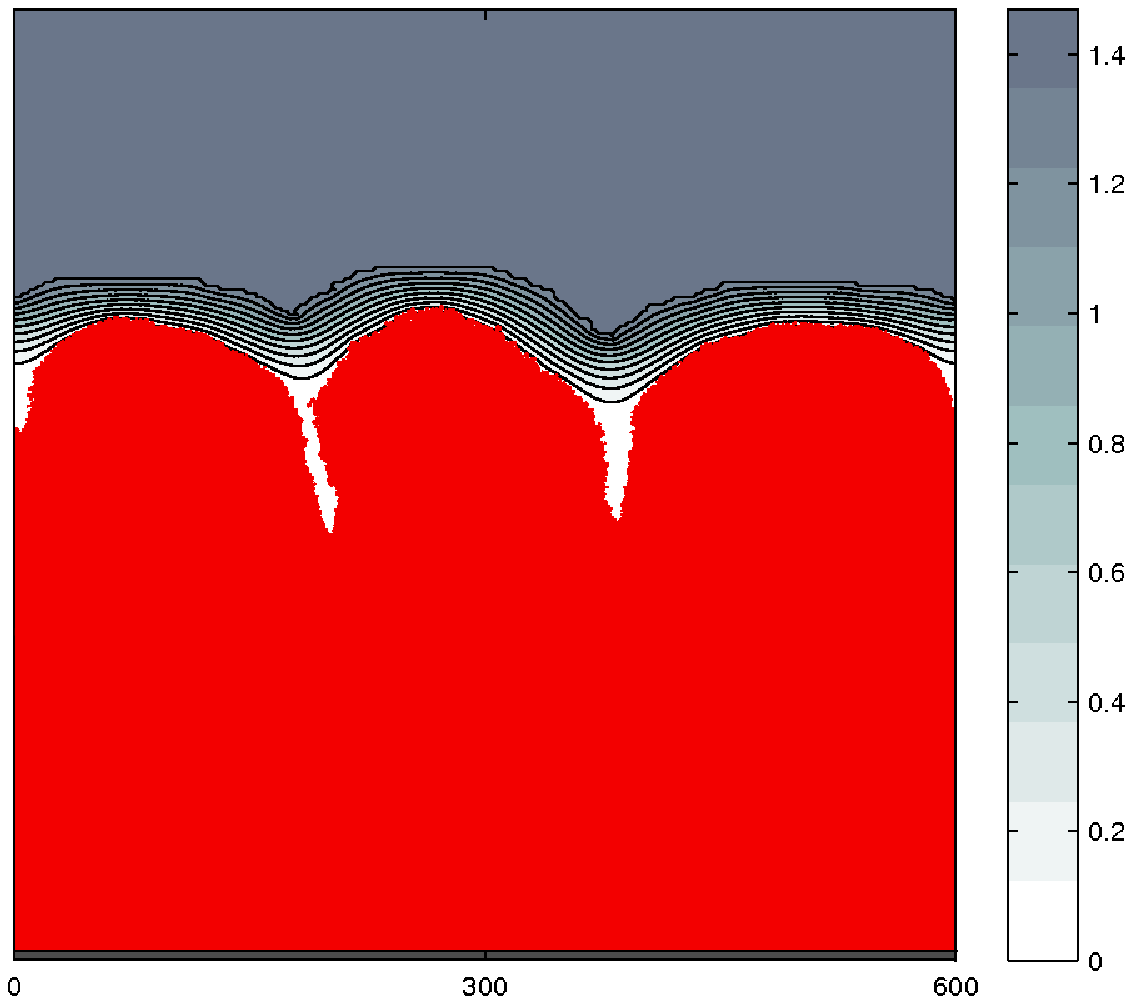}\hspace*{1cm}
\includegraphics[width=7cm]{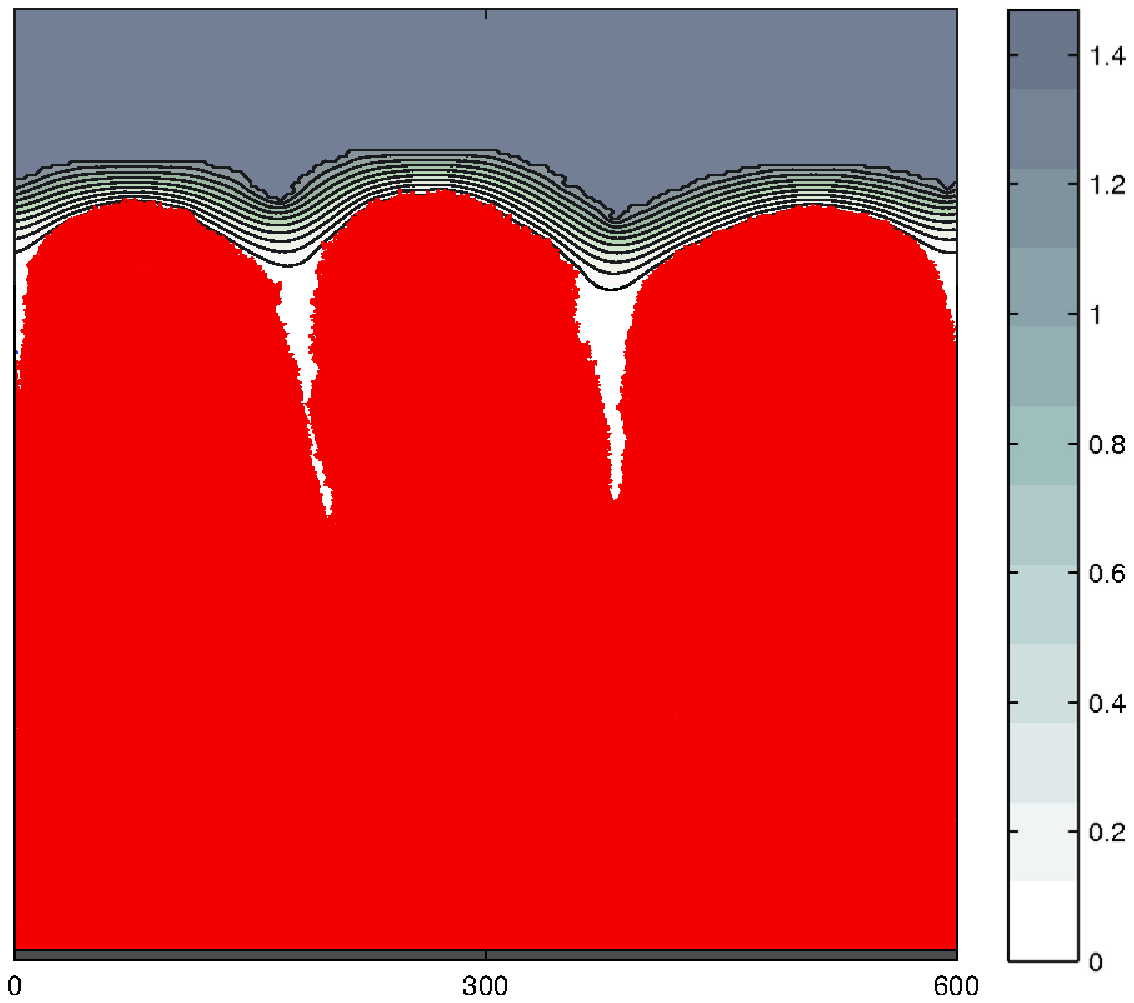}
\caption{\footnotesize{Generation of effective pinning sites by means of the feedback mechanism described in the text. The nutrient concentration at the bottom of the valleys is so low that cells stop growing. Cells at the top, however, can access the bulk concentration, growing at a much larger rate (see \cite{note2}). Left: State of the colony at time $t=41$h. Right: state of the colony $8$ hours later.}}
\label{qsites}
\end{center}
\end{figure}

\vspace{0.25cm}
\noindent
In our case, quenchedness is an emergent feature of the colony. It is the result of a non-trivial interaction between cells and the nutrient field gradient. The growth of each individual is entirely dependent on the amount of nutrient that it can access (see above), which is conditional on the interaction between the nutrient field and the focal cell and its neighbors. Individuals residing at the peak of a region in the colony with larger height than its immediate surroundings have better access to nutrients and grow more quickly. On the other hand, growth rate is reduced among individuals at the bottom of regions with smaller height than the surroundings, where nutrient concentration is reduced due to consumption by neighboring cells; the propagation of the colony front is therefore reduced in these ``valleys'' (see Fig.\ref{qsites}).

\vspace{0.25cm}
\noindent
Competition between cells for the available resource can thus generate points at the interface with substantially reduced propagation velocity \cite{pinning_ex}. Moreover, this inequality of propagation is subject to positive feedback: the slower the propagation of the front at a site in comparison with the neighbors, the deeper the valley becomes and the more severely nutrient access decreases. This feedback effect can cause individuals in the valley to stop growing entirely, creating effective pinning sites along the front. Such pinning sites can only be overcome by the interface by means of an increased external pulling force (nutrient concentration, $C$, in this case) or by lateral growth, which are the same mechanisms exhibited by qKPZ interfaces. 

\vspace{0.25cm}
\noindent
For the values of bulk nutrient concentration used for our measurements (see above), the pinning sites are always eventually overcome, behavior that corresponds to the moving phase of a propagating front advancing through a random environment (quenched disorder). In this way, we can see that the bacterial profiles of our agent-based model have the same basic ingredients and quantitative behavior of a qKPZ interface in its supercritical phase. For system sizes similar to those of our framework, the theoretical finite-size exponents of that class agree with our measurements (see Table 1).

\subsubsection*{\it A Richer Rough Phase}{}

\begin{figure}[t]
\begin{center}
\includegraphics[width=8cm]{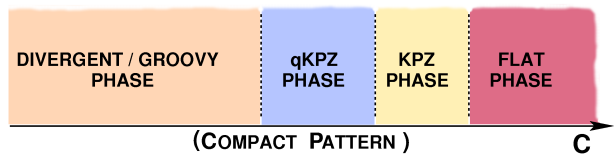}
\caption{\footnotesize{Sketch of the rough part (zone V) of the phase diagram, showing the different sub-phases discussed in the text.}}
\label{arrow}
\end{center}
\end{figure}

In addition to the quenched noise, we must consider the stochasticity inherent in the shoving events among growing and dividing cells for a full description of cell colony patterns. We will aim to explain how the competition between these two noises could act as the nutrient concentration is varied in the rough regime, by using the general equation:

\begin{equation}
\partial_{t}h(x,t)=\lambda(\nabla h)^{2}+D\nabla^{2}h+F+\sigma_{1}\eta_{1}(x,t)+\sigma_{2}\eta_{2}(x,h),
\label{qKPZ_eq} 
\end{equation}

\noindent
with $F$ and the ratio $\sigma_{1}/\sigma_{2}$ changing with $C$:

\begin{itemize}
\item[\texttwelveudash] For a medium nutrient concentration, the resource is quickly depleted in valleys, while the rest of the front can grow at different velocities. Pinning sites therefore emerge easily, while the number of shoving events is low because the growth rate and, thus, the number of birth events, depend on nutrient concentration \cite{note2}. In Eq.(\ref{qKPZ_eq}), this is translated into a small $\sigma_{1}/\sigma_{2}$. Therefore, this source of annealed noise remains asymptotically irrelevant for the macroscopic behavior of the front as compared with the {\it self-generated quenched noise}. It is the reported moving-qKPZ regime.
\item[\texttwelveudash] As nutrient concentration is increased, the growth rate of the cells also increases and so does the mean velocity of the interface. Birth events and the subsequent shoving are more frequent. This hinders the formation of deep valleys, eventually overcome by the advancing colony front. This corresponds to both a larger $F$ and a larger amount of annealed noise (larger $\sigma_{1}$), leading to a behavior that would be governed by the KPZ equation due to the irrelevance of the present quenchedness. 
\item[\texttwelveudash] If the bulk nutrient concentration is further increased, most or all cells in the colony layer have access to inexhaustible resources and grow at the maximum rate at all times ($F\gg1$), leading to a completely flat colony surface. A summary sketch can be found in Fig.\ref{arrow}.
\end{itemize}

\vspace{0.25cm}
These observations motivate further exploration of the parameter space, varying the value of the nutrient concentration in search for the different regimes described above. While we could find a range of concentrations with exponents in agreement with those of qKPZ universality class (see Table 1), we were not able to measure, for any $C$, a set of exponents fully compatible with those of the KPZ class. This might be due to the time limitation of the framework: the higher the value of bulk nutrient concentration, the shorter the time required by the growing colony to reach $h_{max}$ (see above), making it more difficult to look for crossovers in high $C$ regimes. We may, thus, attribute the missed KPZ regime to a lack of observation of the asymptotic behavior of the observables. We are currently developing a new version of the simulation framework in order to overcome this technical limitation \cite{next_bac}.

\vspace{0.25cm}
\noindent
At very high nutrient concentrations, a completely flat interface is easily recovered. The transition from a flat behavior of $W(L,t)$ to a non-trivial value for its associated exponents indicates that the bacterial colony undergoes a roughening phase transition. However, such flat examples will be hardy seen under real experimental conditions (see below).

\vspace{0.25cm}
At very low bulk nutrient concentration the interface cannot overcome pinning sites, while the rest of the active layer continues growing. This behavior is shown in Fig.\ref{qsites}: it does not correspond to the supercritical phase of a qKPZ system, nor its subcritical phase, because the interface never becomes completely inactive. For such low nutrient concentrations, the equivalence with surfaces propagating in random environments is not valid anymore. In this regime, the large differences in velocity between the growing parts of the front and the pinned sites make the $\beta$ exponent take unusually large values around $\beta=1$ (with patterns similar to those reported for the ``groovy'' phase in \cite{fungi1}). At even smaller bulk nutrient concentrations, we enter a phase in which wide pinning regions are developed, but the colony continues to grow at a slow pace. Here, branches arise and the colony can grow not only upwards and laterally, but also downwards due to the presence of more ample free space. This description corresponds to patterns that belong to zone IV in Fig.\ref{orig_phase_diagram}.

\section*{Conclusions}

Bacterial colonies are among the best examples of biological systems whose complexity can be informatively reduced by using concepts from physics. Here, we have shown that the essential biological and physical ingredients shaping colony propagating fronts may be distinguished by quantifying the scaling behavior of a colony.

\vspace{0.25cm}
\noindent
By using a detailed individual-based model in lieu of real experiments, we have shown that cell shape and substratum heterogeneities \texttwelveudash\; in contrast to the existing literature \texttwelveudash\; are not necessary to explain the form and behavior of colony fronts in the rough regime. We have also hypothesized that this regime contains a much richer morphological variety than expected. Based on the biological and physical ingredients present in our simulation framework, we have given biological justifications for the emergence of quenched disorder as the result of competition for space and resources among cells. We have further described how higher nutrient concentration allows colonies to overcome pinning sites due to higher cell growth rates, giving rise to different universal behaviors. Overcoming the technical limitations of the simulation framework will grant better access to the asymptotic behavior of the front in each of these sub-phases.

\vspace{0.25cm}
Classifying the universal behavior of a generic bacterial colony is more than a mere academic exercise. It allows us to describe the behavior of the front by means of one single equation, such as Eq.(\ref{qKPZ_eq}), instead of using explicit agent-based or reaction-diffusion theoretical models. We have observed that irregular cell shape, long-range cell motility, and extracellular compounds are not necessary for our idealized cell groups to exhibit the same universal behavior as real bacterial colonies. This result suggests that such factors may not be responsible for the behavior of the colony patterns observed in experiments, and that the same universal behavior may be expected for the morphology of any bacterial colony under similar conditions.

\vspace{0.25cm}
However, for the moment, we do not have enough information to prove this claim. Only experimental work with different strains will allow us to check if the richer universal behavior described here can be observed even in the presence of other ingredients that can be potentially important for the real asymptotic behavior of colony fronts \cite{next_bac}. For instance, due to the average mutation rate of many commonly studied bacteria, mutations that affect growth rate often arise on the time scale reported for colony growth experiments that we have discussed. A rare minority of these mutants exhibits higher growth rates than the rest of the population, resulting in bursting sectors that emerge from growing colonies. These sectors obviously influence the shape of colony fronts in a way that is not considered by any theoretical models in the literature so far. These mutations will prevent also the observation of the roughening transition in real experiments: the larger the value of $C$, the larger the number of replication events and the higher the likelihood of obtaining mutants with a larger growth rate. This could generate, by different means, large lateral correlations leading to a large $\alpha_{exp}$ such as that measured in the cited experimental studies.

\vspace{0.25cm}
All of the open questions discussed here highlight the necessity of a complete quantitative characterization of the propagating of real bacterial colonies in the rough regime. With it, the biology of the interaction between individuals and the physics of the environment, studied from a multidisciplinary point of view, will allow us to discern the relevant mechanisms responsible for the asymptotic behavior of real bacterial colony profiles.

\section*{Acknowledgments}
We would like to thank M.A. Mu\~noz, for helpful discussions and suggestions. We gratefully acknowledge support from the Defense Advanced Research Projects Agency (DARPA) under grants
HR0011-05-1-0057 and HR0011-09-1-055. CDN is supported by a Princeton University Centennial Fellowship and an NSF graduate research fellowship.

\begin{small}
\bibliography{bacteria.bib}

\begin{thebibliography}{10}
\bibitem{note2}
In our simulations it is assumed a Monod-like relation between the growth rate
  and the nutrient concentration \cite{monod}:
  $\mu=\mu_{max}\frac{C_{loc}}{C_{loc}+K}$, where $\mu_{max}$ is the maximum
  growth rate and $K$ stands for the half-saturation constant, that measures
  the ability of the cell for nutrient uptake.
\bibitem{note}
It is possible to rewrite the Family-Vicsek scaling equations,
  Eq.(\ref{family}), by using universal functions:
  $W(L,t)=t^{\beta}\mathcal{F}\left(L/t^{1/z}\right)$, with $\mathcal{F}(x)\sim
  \textnormal{constant}$ for $x\gg1$ and $\mathcal{F}(x)\sim x^{\alpha}$ for
  $x\ll1$; or $W(L,t)=L^{\alpha}\mathcal{G}\left(t/L^{z}\right)$, with
  $\mathcal{G}(x)\sim \textnormal{constant}$ for $x\gg1$ and
  $\mathcal{G}(x)\sim x^{\beta}$ for $x\ll1$.
\bibitem{note4}
One illustrative example of it is the reaction-diffusion model introduced in
  \cite{bac_model3}. If $A(x,y;t)$ is the field representing the active cells,
  $I(x,y;t)$ the inactive ones, and $n(x,y;t)$ the nutrient field, the dynamics
  is given by $\partial_{t}A=\nabla\left(D(A,n)\nabla A\right)+nA-g(n)A$,
  $\partial_{t}I=g(n)A$, and $\partial_{t}n=\nabla^{2}n-nA$. The rough patterns
  are recovered by choosing $D(A,n)=D_{1}A$, with $D_{1}$ constant. This
  changes the first term in the r.h.s. of the dynamic equation for $A$ to
  $\nabla^{2}A+\left(\nabla A\right)^{2}$. The height of the interface at
  $(x,t)$ corresponds to the highest position $y$ for which $A(x,y;t)\ne0$
  \cite{KPZ_MSM}.
\end{thebibliography}
\bibliographystyle{unsrt}
\end{small}

\end{document}